\documentclass[graybox]{svmult}

\usepackage{type1cm}        
\usepackage{makeidx}         
\usepackage{graphicx}        
\usepackage{multicol}        
\usepackage[bottom]{footmisc}

\usepackage{newtxtext}       %
\usepackage{newtxmath}       

\makeindex             

\def\x{{\bf x}}
\def\X{{\bf X}}
\def\IQR{{\rm IQR}}
\def\UQR{{\rm UQR}}
\def\LQR{{\rm LQR}}
\def\MAD{{\rm MAD}}
\def\med{{\rm med}}
\def\R{\ifmmode{I\hskip -3pt R}
           \else{\hbox{$I\hskip -3pt R$}}\fi}

\begin{document}

\title*{Minkowski distances and standardisation for clustering and classification on high dimensional data}
\titlerunning{Distances for high dimensional data} 
\author{Christian Hennig}
\institute{Christian Hennig \at Dipartimento di Scienze Statistiche ``Paolo Fortunati'', Via delle Belle Arti 41, 40126 Bologna, Italy  \email{christian.hennig@unibo.it}}
%
%
\maketitle

\abstract{There are many distance-based methods for classification and clustering, and for data with a high number of dimensions and a lower number of observations, processing distances is computationally advantageous compared to the raw data matrix. Euclidean distances are used as a default for continuous multivariate data, but there are alternatives. Here $L_1$ (city block)-, $L_2$ (Euclidean)-, $L_3$, $L_4$-, and maximum distances are combined with different schemes of standardisation of the variables before aggregating them. Boxplot transformation, a new transformation method for a single variable that standardises the majority of observations but brings outliers closer to the main bulk is proposed. Distances are compared in simulations for clustering by partitioning around medoids, complete and average Linkage, and classification by nearest neighbours, of data with low number of observations but high dimensionality. The $L_1$-distance and the boxplot transformation show good results.}  

\section{Introduction}
\label{sec:1}
{\it The final version of this has now been published:\\
Hennig, C.: Minkowski distances and standardisation for clustering and classification on high dimensional data. In: Imaizumi, Tadashi, Nakayama, Atsuho, Yokoyama, Satoru (Eds.) ``Advanced Studies in Behaviormetrics and Data Science.
Essays in Honor of Akinori Okada'', Springer Singapore (2020), p. 103-118.}
~\\~\\ 
One thing that I share with Professor Akinori Okada is the affinity for dissimilarities and distances. At the IFCS 2017 in Tokyo, when Professor Okada was President of the International Federation of Classification Societies and I was Secretary, he gave a fascinating presentation on ``dissimilarity based on dissimilarity to others'', and his work is full of nonstandard takes on dissimilarity. In the present paper the dissimilarities of interest will be fairly standard and symmetric distances, but they will still be used in ways that are, in my opionion, underrepresented and underinvestigated in the literature.  

There are many dissimilarity-based methods for clustering and supervised classification, for example partitioning around medoids, the classical hierarchical linkage methods (\cite{KauRou90}) and $k$-nearest neighbours classification (\cite{CovHar67}). Approaches such as multidimensional scaling are also based on dissimilarity data. There is much literature on the construction and choice of dissimilarities (or, mostly equivalently, similarities) for various kinds of nonstandard data such as images, melodies, or mixed type data. For standard quantitative data, however, non-dissimilarity analysis is often preferred (some of which implicitly rely on the Euclidean distance, particularly when based on Gaussian distributions), and where  dissimilarity-based methods are used, in most cases the Euclidean distance is employed. In the following,  all considered dissimilarities will fulfill the triangle inequality and therefore be distances. 

Given a data matrix of $n$ observations in $p$ dimensions $\X=(\x_1,\ldots,x_n)$ where $\x_i=(x_{i1},\ldots,x_{ip})\in\R^p,\ i=1,\ldots,n$, in case that $p>n$, analysis of $(n-1)(n-2)/2$ distances $d(\x_i,\x_j)$ is computationally advantageous compared with the analysis of $np$ raw data matrix entries. High dimensionality comes with a number of issues (often referred to as the ``curse of dimensionality''; e.g., \cite{HMN05}). This could make distances attractive for high dimensional data, particularly because the distances do not directly carry information about the dimensionality of the data. But some issues with high dimensions manifest themselves also at the level of distances, for example some at first sight fairly innocent distributional conditions allow to prove that for fixed $n$ and $p\to\infty$ all Euclidean distances between points converge to a constant (\cite{AMMC07}). There are however indications that not all reasonable distances are affected in the same way by such problems (\cite{HMN05}). \cite{Murtagh09} takes a different point of view and argues that the structure of very high dimensional data can even be advantageous for clustering, because distances tend to be closer to ultrametrics, which are fitted by hierarchical clustering. He also demonstrates that the components of mixtures of separated Gaussian distributions can be well distinguished in high dimensions, despite the general tendency toward a constant. Similarly, for classification \cite{AMMC07} state that class separation corresponding to the underlying data generation process is still possible. 

Here I investigate a number of distances when used for clustering and supervised classification for data with low $n$ and highy $p$, with a focus on two ingredients of distance construction, for which there are various possibilities, namely {\it standardisation}, i.e., some usually linear transformation based on variation in order to make variables with differing variation comparable, and {\it aggregation} of a single distance out of the contributions of the individual variables. Particular attention is devoted to the impact of outliers. A new proposal for standardisation, the boxplot transformation, which does not only standardise variables unaffected by outliers, but also brings such outliers closer to the main bulk of the data.  

In Section \ref{sconstruction}, besondes some general discussion on distance construction, various proposals for two building blocks of a distance, namely standardisation of individual variables and aggregation, are made. Section \ref{sexp} presents a simulation study comparing the different combinations of standardisation and aggregation. Section \ref{sconclusion} concludes the paper.   

\section{Distance construction}
\label{sconstruction}
The distances considered here are constructed as follows. First, the variables are standardised in order to make them suitable for aggregation, then they are aggregated according to Minkowski's $L_q$-principle. These two steps can be found often in the lterature, however their joint impact and performance for high dimensional classification has hardly been investigated systematically. Before introducing the standardisation and aggregation methods to be compared, the section is opened by a discussion of the differences between clustering and supervised classification problems. 
\subsection{Clustering vs. supervised classification}
\label{sclussuper}
Superficially, clustering and supervised classification seem very similar. A popular assumption is that for the data there exist true class labels $C_1,\ldots,C_n\in \{1,\ldots,k\}$, and the task is to estimate them. In clustering, all $C_1,\ldots,C_n$ are unknown, whereas in supervised classification they are known, and the task is to construct a classificaton rule to classify new observations, i.e., to estimate $C_{n+1},\ldots,C_{n+m}\in \{1,\ldots,k\}$ for given $\x_{n+1},\ldots,\x_{n+m}$. Obviously the clustering problem is more difficult due to less available information, but underlying model assumptions could be taken to be the same, and approaches could be expected to be related. When comparing distances in a simulation in Section \ref{sexp}, this is in fact the approach that will be taken. I have however argued in \cite{Hennig15} that the clustering situation is somewhat more different in the sense that there could be various legitimate ``true'' clusterings on the same dataset, and that it should depend on background knowledge and the aim of clustering what kind of clustering should be preferred. Particularly, decisions made before the application of a clustering method such as standardisation, variable selection, and distance construction, do not only influence the result, but can also be seen as implicitly contributing to the definition of the clusters that will be found. For example, if the data stems from a questionnaire that has general demographic questions as well as detailed questions about a certain issue such as the usage of means of transport, one could be interested in a clustering that aggregates demographic and transport usage information, but one could also be interested in a clustering based on transport usage alone that may or may not be in some way informed by the demographic information. 

An issue regarding standardisation is whether different variances of variables are seen as informative in the sense that a larger variance means that the variable shows a ``signal'', whereas a low variance means that mostly noise is observed. This happens in a number of engineering applications, and in this case standardisation that attempts to making the variances equal should be avoided, because this would remove the information in the variances. If class labels are given, as in supervised classification, it is just possible to compare these alternatives using the estimated misclassification probability from cross-validation and the like. However, in clustering such information is not given. The data therefore cannot decide this issue automatically, and the decision needs to be made from backgound knowledge. It is even conceivable that for some data both use of or refraining from standardisation can make sense, depending on the aim of clustering. When analysing high dimensional data such as from genetic microarrays, however, there is often not much background knowledge about the individual variables that would allow to make such decisions, so users will often have to rely on knowledge coming from experiments as in Section \ref{sexp} with a single given true clustering.

Lastly, in supervised classification class information can be used for standardisation, so that it is possible, for example to pool within-class variances, which are not available in clustering. 
\subsection{Standardisation}
Normally, and for all methods proposed in Section \ref{sagg}, aggregation of information from different variables in a single distance assumes that ``local distances'', i.e., differences between observations on the individual variables, can be meaningfully compared. This is obviously not the case if the variables have incompatible measurement units, and fairly generally more variation will give a variable more influence on the aggregated distance, which is often not desirable (but see the discussion in Section \ref{sclussuper}). None of the aggregation methods in Section \ref{sagg} is scale invariant, i.e., multiplying the values of different variables with different constants (e.g., changes of measurment units) will affect the results of distance-based clustering and supervised classification. Therefore standardisation in order to make local distances on individual variables comparable is an essential step in distance construction.

Normally, standardisation is carried out as
$$
x_{ij}^*=\frac{x_{ij}-a_j^*}{s_j^*},\ i=1,\ldots,n,\ j=1,\ldots,p,
$$
where $a_j^*$ is a location statistic and $s_j^*$ is a scale statistic depending on the data. For distances based on differences on individual variables as used here, $a_j^*$ can be ignored, because it does not have an impact on differences between two values. 

The most popular standardisation is standardisation to unit variance, for which $(s_j^*)^2=s_j^2=\frac{1}{n-1}\sum_{i=1}^n(x_{ij}-a_j)^2$ with $a_j$ being the mean of variable $j$. The sample variance $s_j^2$ can be heavily influenced by outliers, though, and therefore in robust statistics often the median absolute deviation from the median (MAD) is used, $s_j^*=\MAD_j=\med\left|\left(x_{ij}-\med_{j}(\X)\right)_{i=1,\ldots,n}\right|$ (by $\med_{j}$ I denote the median of variable $j$ in dataset $\X$, analogously later $\min_j$ and $\max_j$). The ``outliers'' to be negotiated here are outlying values on single variables, and their effects on the aggregated distance involving the observation here they occur; this is not about full outlying $p$-dimensional observations (as are often treated in robust statistics). In high dimensional data often all or almost all observations are affected by outliers in some variables. 

If standardisation is used for distance construction, this does not necessarily solve the issue of outliers. If the MAD is used, the variation of the different variables is measured in a way unaffected by outliers, but the outliers are still in the data. Unit variance standardisation may undesirably reduce the influence of the non-outliers on a variable with gross outliers, which does not happen with MAD-standardisation, but after MAD-standardisation a gross outlier on a standardised variable can still be a gross outlier and may dominate the influence of the other variables when aggregating them. One aim of the simulation in Section \ref{sexp} is to compare the impact of these two issues.

A third approach to standardisation is standardisation to unit range, with 
$s_j^*=r_{j}=\max_j(\X)-\min_j(\X)$. This is influenced even stronger by extreme observations than the variance. But \cite{MilCoo88} have observed that range standardisation is often superior for clustering, namely in case that a large variance (or MAD) is caused by large differences between clusters rather than within clusters, which is useful information for clustering and will be weighted down stronger by unit variance or MAD-standardisation than by range standardisation. The same argument holds for supervised classification.

For this reason it can be expected that a better standardisation can be achieved for supervised classification if within-class variances or MADs are used instead of involving between-class differences in the computation of the scale functional. As discussed earlier, this is not available for clustering (but see \cite{ArGnKe82}, who pool variances within estimated clusters in an iterative fashion). For within-class variances $s_{lj}^2,\ l=1,\ldots,k,\ j=1,\ldots,p$ the pooled within-class variance of variable $j$ is defined as $s_j^*=(s_j^{pool})^2=\frac{1}{\sum_{l=1}^k (n_l-1)}\sum_{l=1}^k (n_l-1)s_{lj}^2,$ where $n_l$ is the number of observations in class $l$. Similarly, with within-class MADs and within-class ranges $\MAD_{lj}, r_{lj},\ l=1,\ldots,k,\ j=1,\ldots,p$, respectively, the pooled within-class MAD of variable $j$ can be defined as $\MAD_j^{poolw}=\frac{1}{n}\sum_{l=1}^k n_l\MAD_{lj}$, and the pooled range as $r_j^{poolw}=\frac{1}{n}\sum_{l=1}^k n_lr_{lj}$ (``weights-based pooled MAD and range''). 

There is an alternative way of defining a pooled MAD by first shifting all classes to the same median and then computing the MAD for the resulting sample (which is then equal to the median of the absolute values) ``shift-based pooled MAD''). For the variance, this way of pooling is equivalent to computing $(s_j^{pool})^2$, because variances are defined by summing up squared distances of all observations to the class means. For the MAD, however, the result will often differ from weights-based pooling, because different observations may end up in the smaller and larger half of values for computing the involved medians. For variable $j=1,\ldots,p$:
$s_j^*=\MAD_j^{pools}=\med_{j}(\X^+)$, where $\X^+=\left(\left|x_{ij}^+\right|\right)_{i=1,\ldots,n,\ j=1,\ldots,p}$, $x_{ij}^+=x_{ij}-\med\left((x_{hj})_{h:\ C_h=C_i}\right)$.  
The same idea applied to the range would mean that all data are shifted so that they are within the same range, which then needs to be the maximum of the ranges of the individual classes $r_{lj}$, so $s_j^*=r_j^{pools}=\max_l r_{lj}$ (``shift-based pooled range''). Whereas in weights-based pooling the classes contribute with weights according to their sizes, shift-based pooling can be dominated by a single class. The shift-based pooled range is determined by the class with the largest range, and the shift-based pooled MAD can be dominated by the class with the smallest MAD, at least if there are enough shifted observations of the other classes within its range.  

\subsection{Boxplot transformation}

\begin{figure}[tb]
\begin{center}
\includegraphics[width=0.6\textwidth]{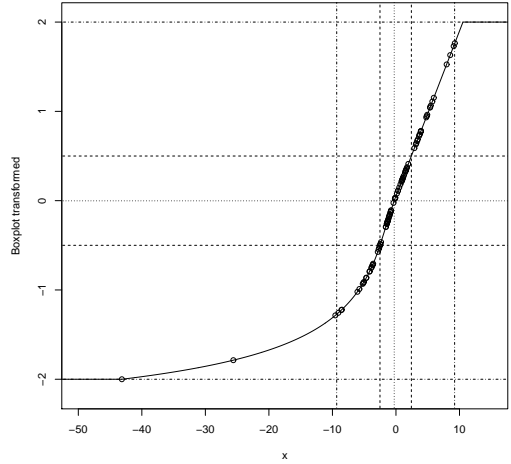}
\end{center}
%
%
\caption{Illustration of boxplot transformation. Scatterplot of dataset $\X$ vs.
boxplot transformed $\X^*$. Lines orthogonal to the $x$-axis are, from left
to right, lower outlier boundary, first quartile, median, third quartile, 
upper outlier boundary. Lines orthogonal to the $y$-axis are, from bottom to top, $-2$ (lower boundary), $-0.5$ (first quartile), median, $0.5$ (third quartile), 2 (upper boundary).}
\label{fboxplotstan}       
\end{figure}

As discussed above, outliers can have a problematic influence on the distance regardless of whether variance, MAD or range is used for standardisation, although their influence plays out differently for these choices. The boxplot standardisation introduced here is meant to tame the influence of outliers on any variable. It is inspired by the outlier identification used in boxplots (\cite{MGTuLa78}). The boxplot shows lower quartile ($q_{1j}(\X)$ where $j=1,\ldots,p$ once more denotes the number of the variable), median ($\med_j(\X)$), and upper quartile ($q_{3j}(\X)$) of the data. It defines as outliers observations for which $x_{ij}<q_{1j}(\X)-1.5\IQR$ or $x_{ij}>q_{3j}(\X)+1.5\IQR_j(\X)$, where $\IQR_j(\X)=q_{3j}(\X)-q_{1j}(\X)$ is the interquartile range. An asymmetric outlier identification more suitable for skew distributions can be defined by using the ranges between the median and the upper and lower quartile, respectively, $\UQR_j(\X)= q_{3j}(\X)-\med_j(\X)$,
$\LQR_j(\X)= \med_j(\X)-q_{1j}(\X)$, so that a lower outlier is defined by $x_{ij}<q_{1j}(\X)-3\LQR_j(\X)$ and an upper outlier by $x_{ij}>q_{3j}(\X)+3\UQR_j(\X)$. The idea of the boxplot transformation is to standardise the lower and upper quantile linearly to $[-0.5,0.5]$. If there are no outliers smaller than the median according to the above rule, all these observations are standardised in the same way, as are all observations larger than the median if there are no outliers. If there are outliers on the lower side of the median, all observations in $[\min_j(\X),q_{1j}(\X)]$ are transformed by a nonlinear transformation that maps the minimum to $-2$ (which is $-0.5-1.5\IQR_j$), so that the outliers are brought so close to the main bulk of the data that they are no longer outliers by the boxplot definition. Analogously, observations in $[q_{3j}(\X),max_j(\X)]$ are mapped to $[0.5,2]$ if there are upper outliers.  
 
The precise computation is as follows. 
\begin{description}
\item[Step 1] Median centering:  
$\X^m=\left(x_{ij}^m\right)_{i=1,\ldots,n,\ j=1,\ldots,p}$ where 
$x_{ij}^m=x_{ij}-\med_j(\X)$.
\item[Step 2] For $j\in\{1,\ldots,p\}$ transform lower quantile to $-0.5$:
For $x_{ij}^m<0:\ x_{ij}^*=\frac{x_{ij}^m}{2\LQR_j(\X^m)}$.
\item[Step 3] For $j\in\{1,\ldots,p\}$ transform upper quantile to 0.5:
For $x_{ij}^m>0:\ x_{ij}^*=\frac{x_{ij}^m}{2\UQR_j(\X^m)}$. 
$\X^*=\left(x_{ij}^*\right)_{i=1,\ldots,n,\ j=1,\ldots,p}.$
\item[Step 4] If there are lower outliers, i.e., $x_{ij}^*<-2$:
  \begin{description}
  \item[Step 4a] Find $t^l_j$ so that $-0.5-\frac{1}{t^l_j}+\frac{1}{t^l_j\left(-\min_j(\X^*)-0.5+1\right)^{t^l_j}}=-2$.
  \item[Step 4b] For $x_{ij}^*<-0.5$: $x_{ij}^*=-0.5-\frac{1}{t^l_j}+\frac{1}{t^l_j\left(-x_{ij}^*-0.5+1\right)^{t^l_j}}.$
  \end{description}
\item[Step 5] If there are upper outliers, i.e., $x_{ij}^*>2$:
  \begin{description}
  \item[Step 5a] Find $t^u_j$ so that $0.5+\frac{1}{t^u_j}-\frac{1}{t^u_j\left(\max_j(\X^*)-0.5+1\right)^{t^u_j}}=2$.
  \item[Step 5b] For $x_{ij}^*>0.5$: $x_{ij}^*=0.5+\frac{1}{t^u_j}-\frac{1}{t^u_j\left(x_{ij}^*-0.5+1\right)^{t^u_j}}.$
  \end{description}
\item[Step 6] In case of supervised classification of new observations, the 
boxplot standardisation is computed as above, using the quantiles, $t^l_j,\ t^u_j$ from the training data $\X$, but values for the new observations are capped to be $\in[-2,2]$, i.e., everything smaller than $-2$ is set to $-2$ and everything larger than 2 is set to 2.  
\end{description}
Figure \ref{fboxplotstan} illustrates the boxplot transformation for a 
given dataset. The boxplot transformation is somewhat similar to a classical technique called Winsorisation (\cite{Ruppert06}) in that it also moves outliers closer to the main bulk of the data, but it is smoother and more flexible.
\subsection{Aggregation}
\label{sagg}
Information from
the variables ais aggregated here by standard Minkowski $L_q$-distances,
$$
d_{Lq}(\x_i,\x_j)=\sqrt[q]{\sum_{l=1}^p d_l(x_{il},x_{jl})^q},
$$
where $q=1$ delivers the so-called city block distance, adding up absolute values of variable-wise differences, $q=2$ corresponds to the Euclidean distance, and $q\to\infty$ will eventually only use the maximum variable-wise difference, sometimes called $L_\infty$ or maximum distance.        

These aggregation schemes treat all variables equally (``impartial aggregation''). Much work on high-dimensional data is based on the paradigm of dimension reduction, i.e., they look for a small set of meaningful dimensions to summarise the information in the data, and on these standard statistical methods can be used hopefully avoiding the curse of dimensionality. Using impartial aggregation, information from all variables is kept. 
There is widespread belief that in many applications in which high-dimensional data arises, the meaningful structure can be found or reproduced in much lower dimensionality. Where this is true, impartial aggregation will keep a lot of high-dimensional noise and is probably inferior to dimension reduction methods. However, there may be cases in which high-dimensional information cannot be reduced so easily, either because meaningful structure is not low dimensional, or because it may be hidden so well that standard dimension reduction approaches do not find it. My impression is that for both dimension reduction and impartial aggregation there are situations in which they are preferable, although they are not compared in the present paper.

A side remark here is that another distance of interest would be the Mahalanobis distance. The Mahalanobis distance is invariant against affine linear transformations of the data, which is much stronger than achieving invariance against changing the scales of individual variables by standardisation. It has been argued that affine equi- and invariance is a central concept in multivariata analysis, see, e.g., \cite{Ruppert06}. But in high dimensions, things change. \cite{PirBra19} show that for $p\ge n-1$ all observations have the same Mahalanobis distance to all other observations. Together with the result of \cite{Tyler10} that for $p\ge n-1$ any affine equivariant scatter statistic must be proportional to the sample covariance matrix as employed in the Mahalanobis distance, it follows that affine invariance cannot be achieved for high dimensional data in a nontrivial manner that is informative about the data. This means that some directions in data space necessarily have to be privileged over others, as are the main coordinate axes for the Minkowski distances.  

\section{Experiments}
\label{sexp}
\begin{figure}[tb]
\begin{center}
\includegraphics[width=0.48\textwidth]{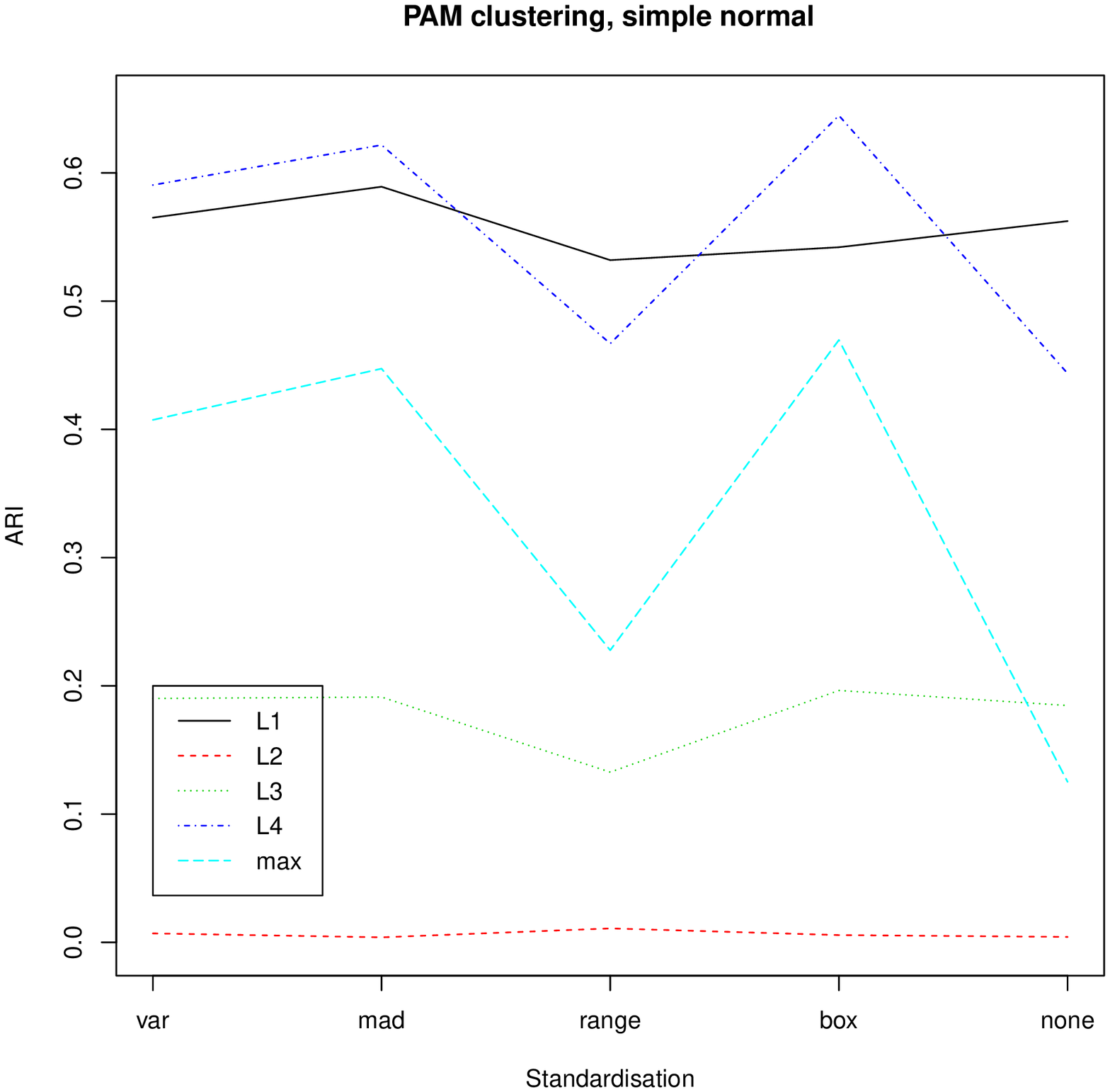}
\includegraphics[width=0.48\textwidth]{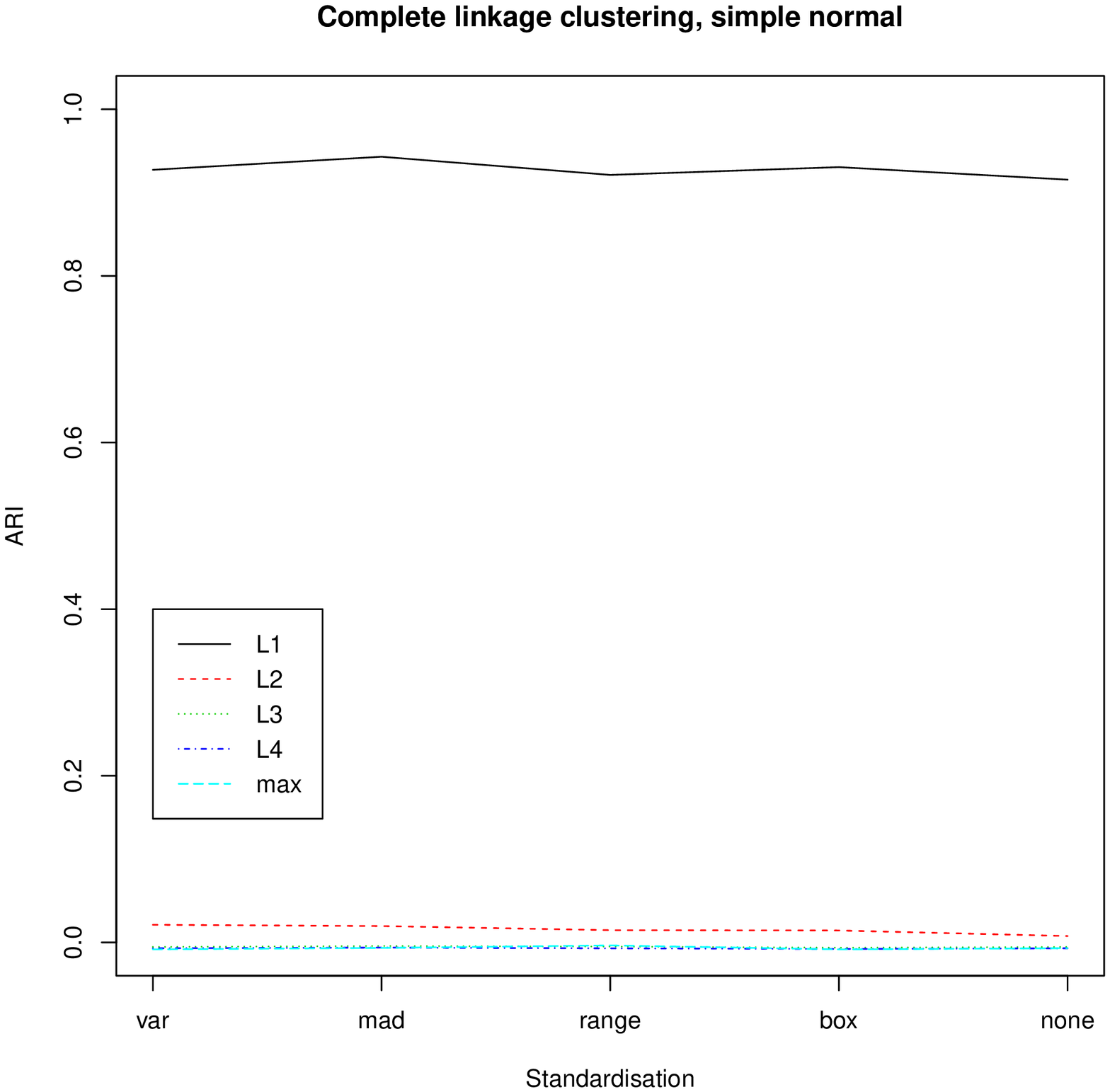}
\includegraphics[width=0.48\textwidth]{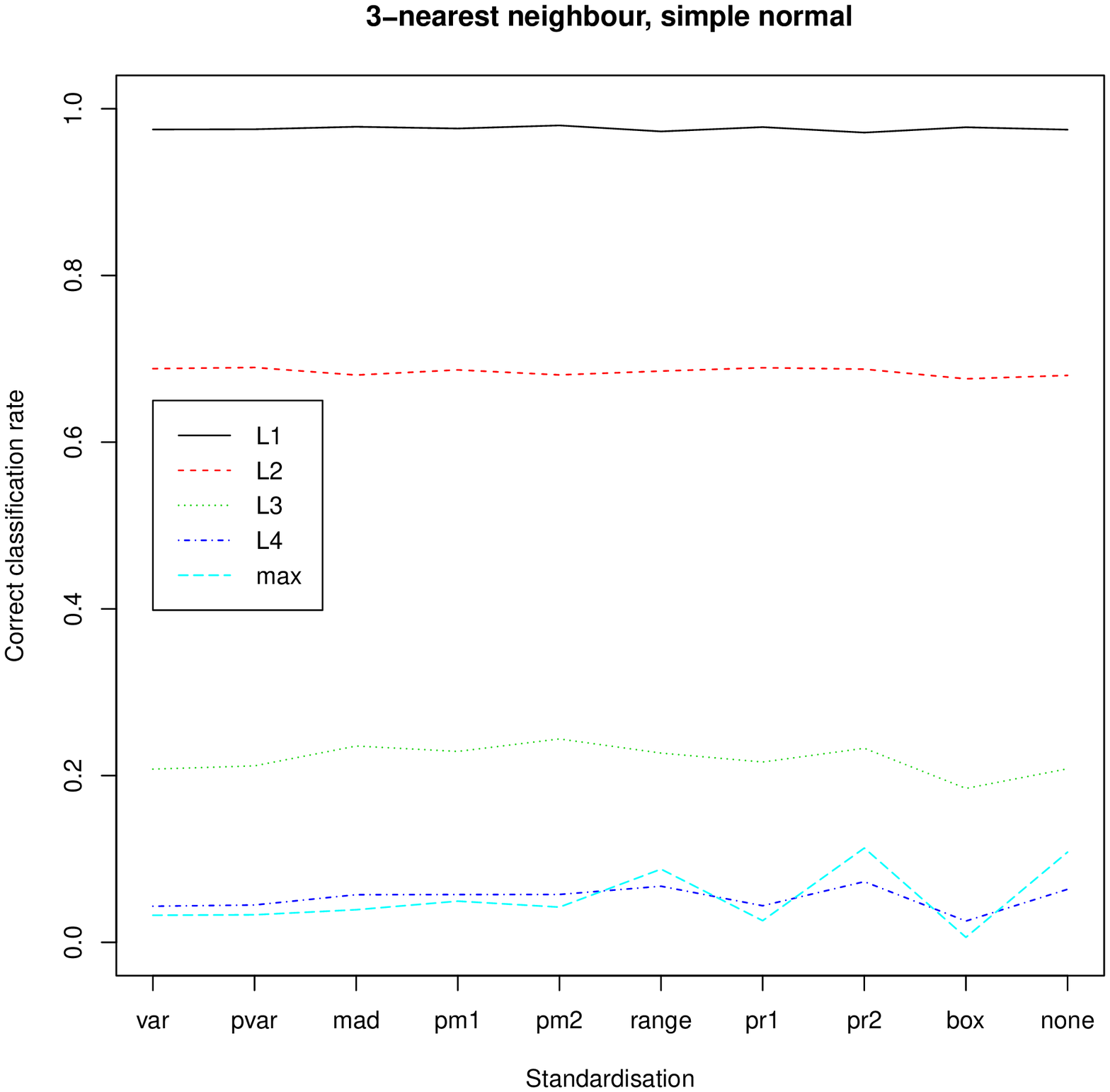}
\end{center}
%
%
\caption{Results from the simple normal setup, adjusted Rand index (ARI) from PAM and complete linkage, and misclassification rates from 3-nearest neighbours.}
\label{fsimplenormal}       
\end{figure}

\begin{figure}[tb]
\begin{center}
\includegraphics[width=0.48\textwidth]{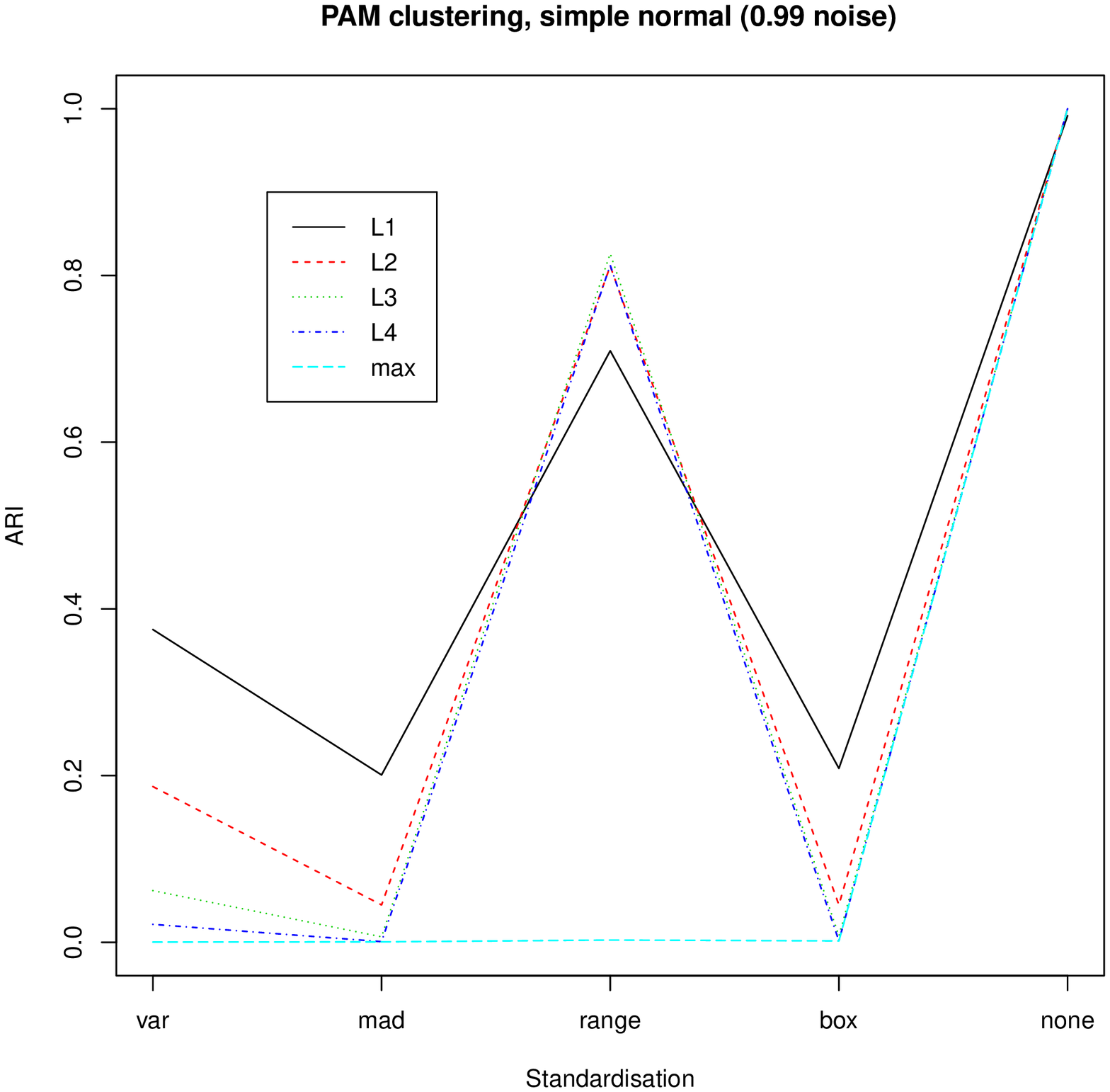}
\includegraphics[width=0.48\textwidth]{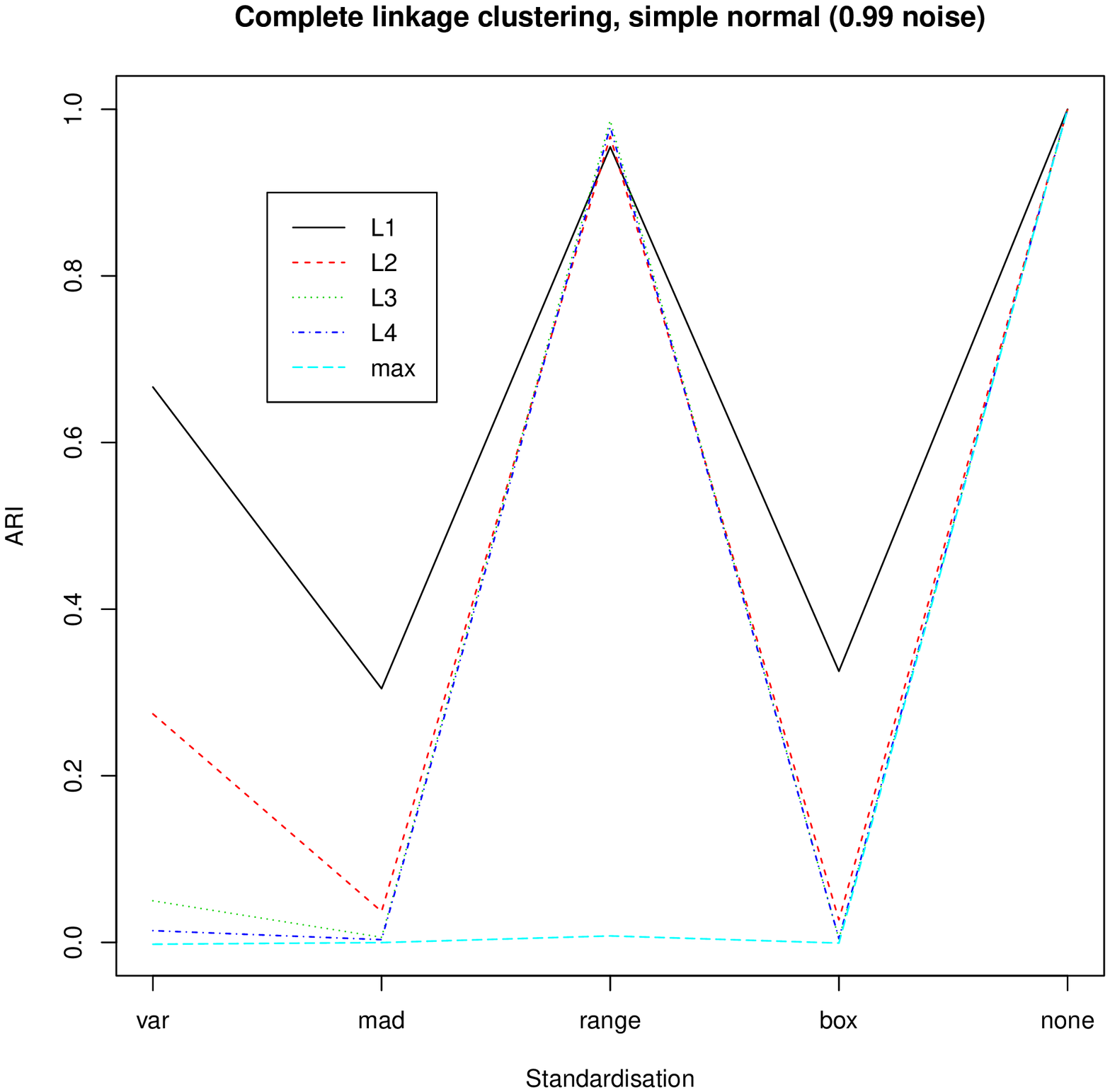}
\includegraphics[width=0.48\textwidth]{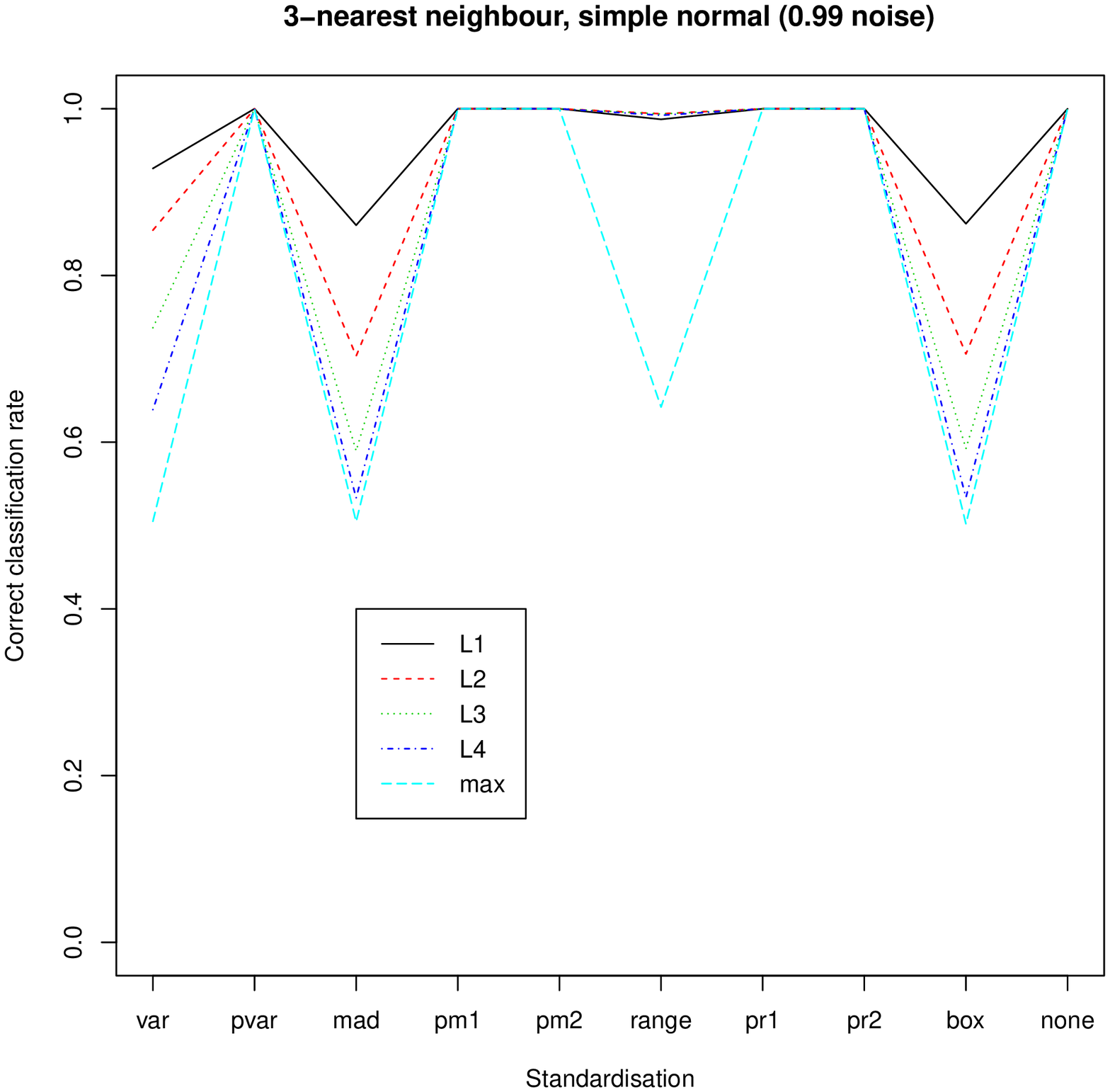}
\end{center}
%
%
\caption{Results from the simple normal (0.99 noise) setup, adjusted Rand index (ARI) from PAM and complete linkage, and misclassification rates from 3-nearest neighbours.}
\label{fsimplenormal099}       
\end{figure}

\begin{figure}[tb]
\begin{center}
\includegraphics[width=0.48\textwidth]{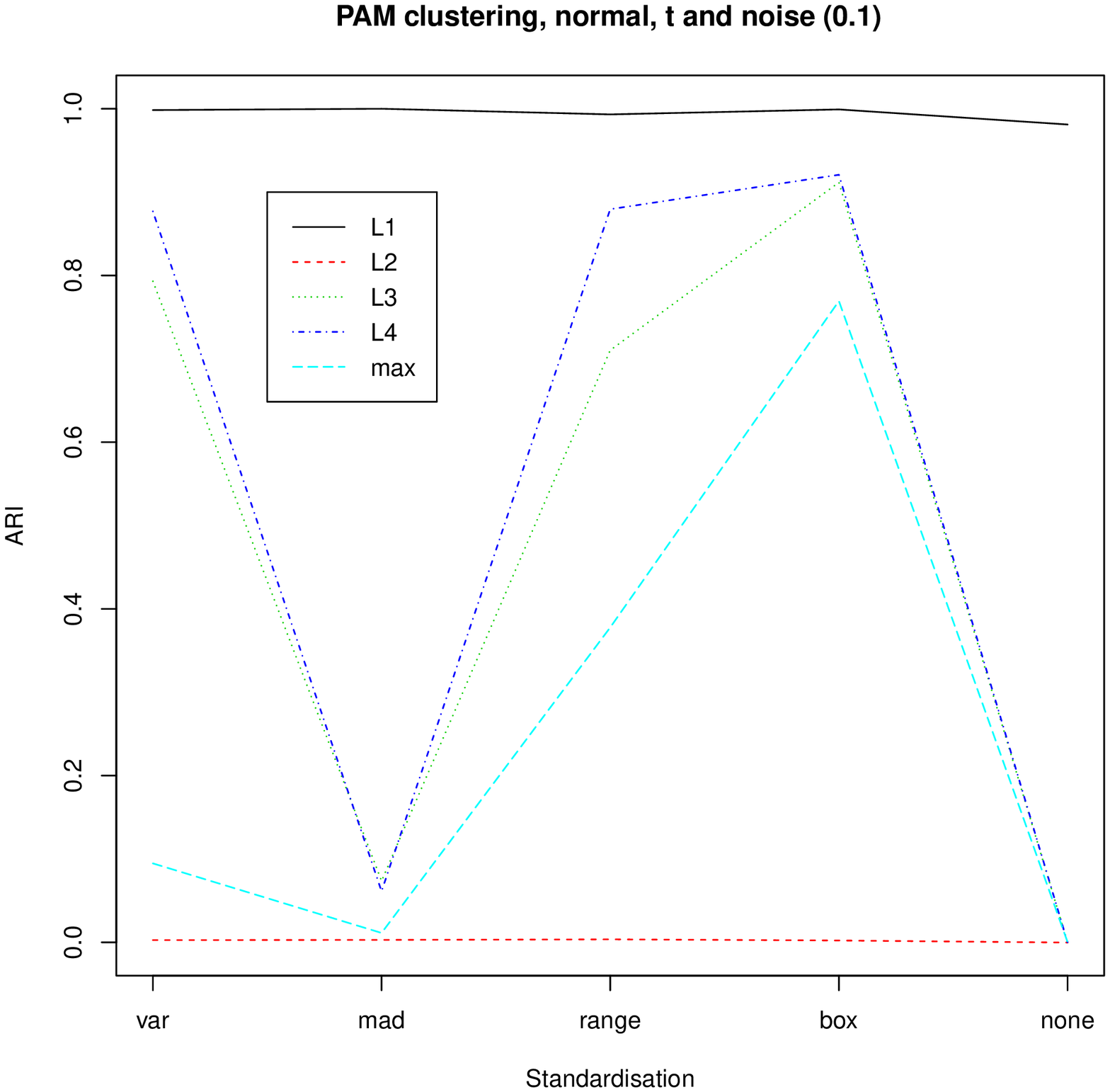}
\includegraphics[width=0.48\textwidth]{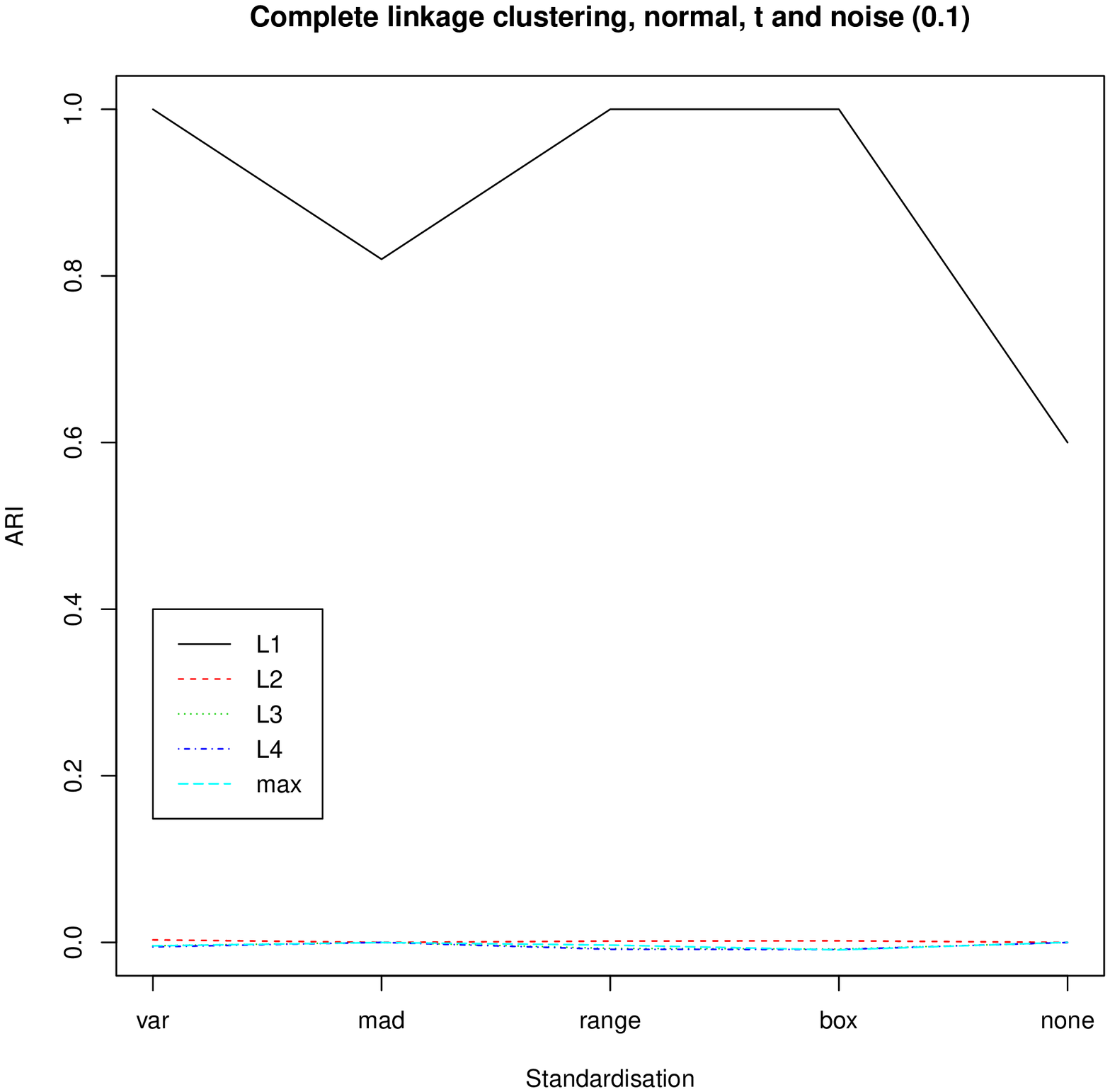}
\includegraphics[width=0.48\textwidth]{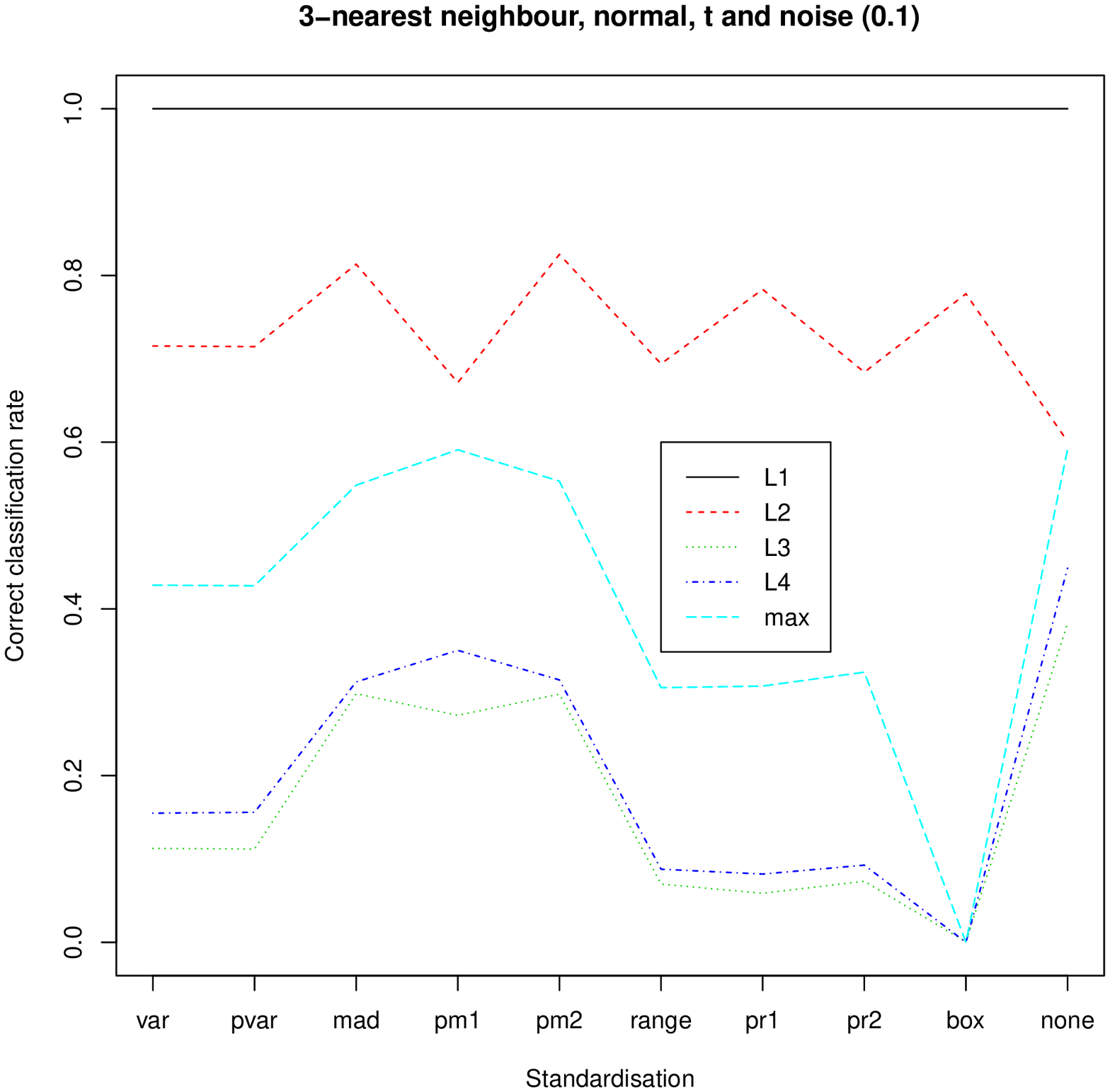}
\end{center}
%
%
\caption{Results from the normal, t, and noise (0.1) setup, adjusted Rand index (ARI) from PAM and complete linkage, and misclassification rates from 3-nearest neighbours.}
\label{fnormaltnoise}       
\end{figure}

\begin{figure}[tb]
\begin{center}
\includegraphics[width=0.48\textwidth]{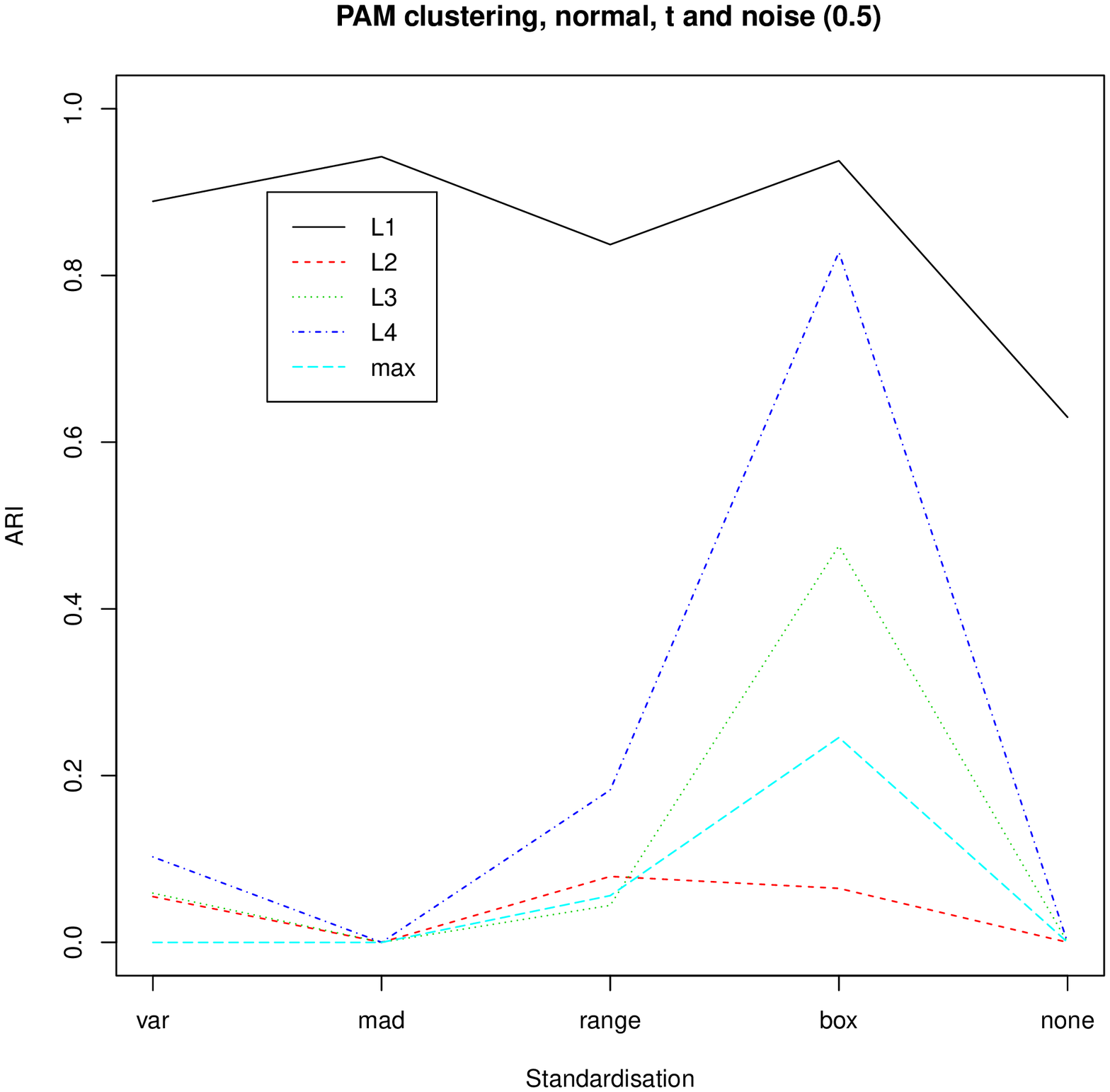}
\includegraphics[width=0.48\textwidth]{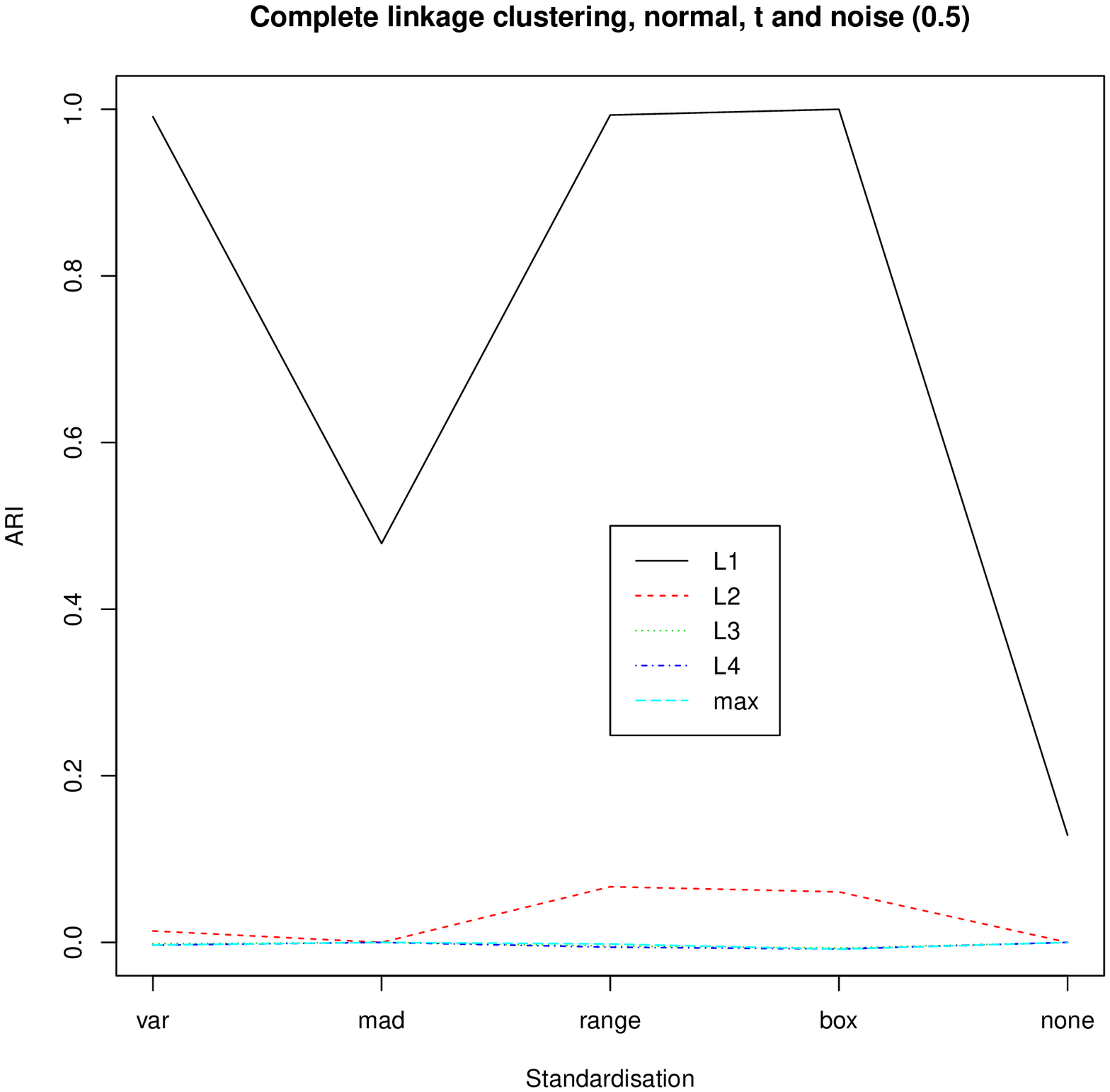}
\includegraphics[width=0.48\textwidth]{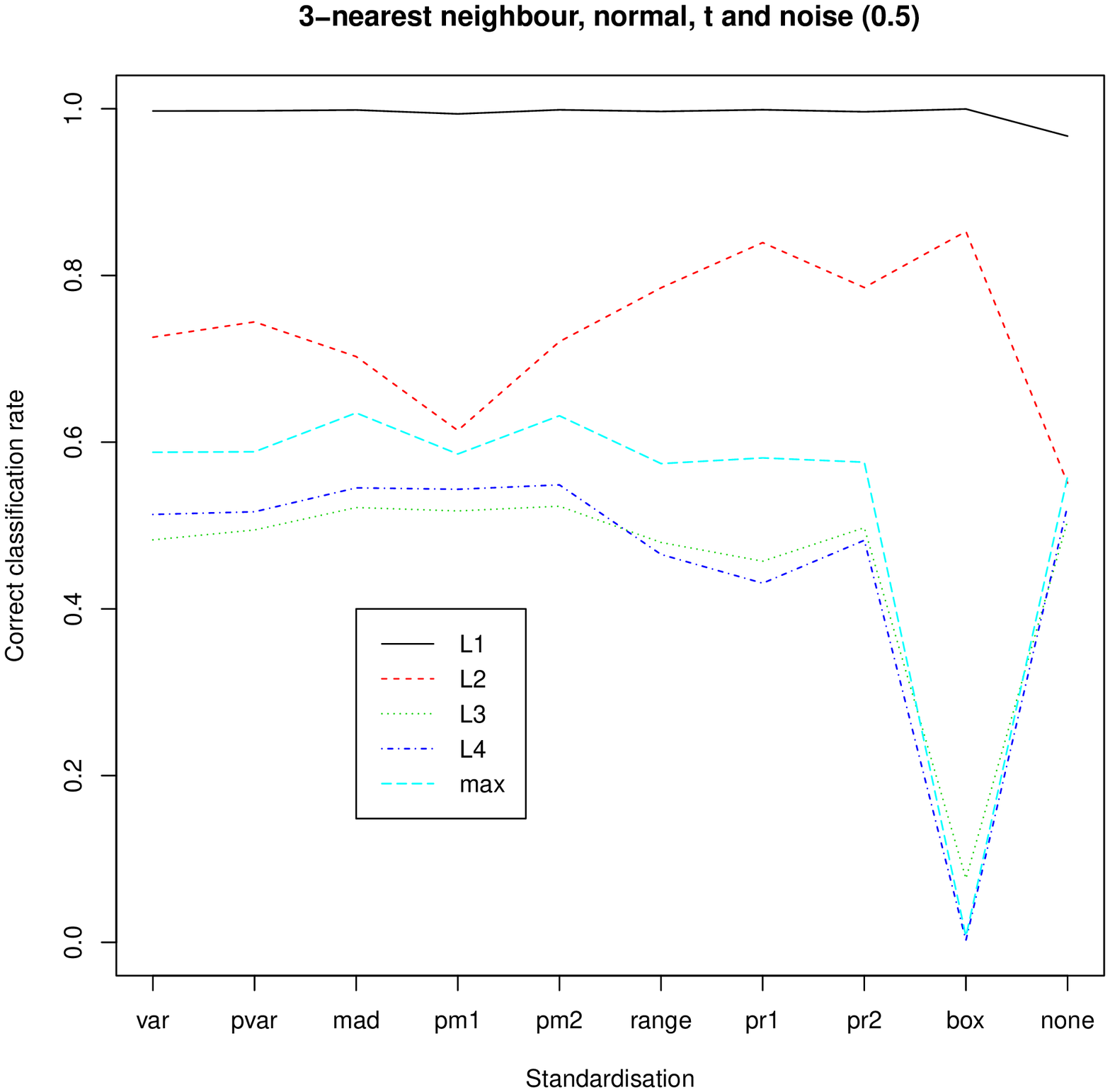}
\end{center}
%
%
\caption{Results from the normal, t, and noise (0.5) setup, adjusted Rand index (ARI) from PAM and complete linkage, and misclassification rates from 3-nearest neighbours.}
\label{fnormaltnoise05}       
\end{figure}

\begin{figure}[tb]
\begin{center}
\includegraphics[width=0.48\textwidth]{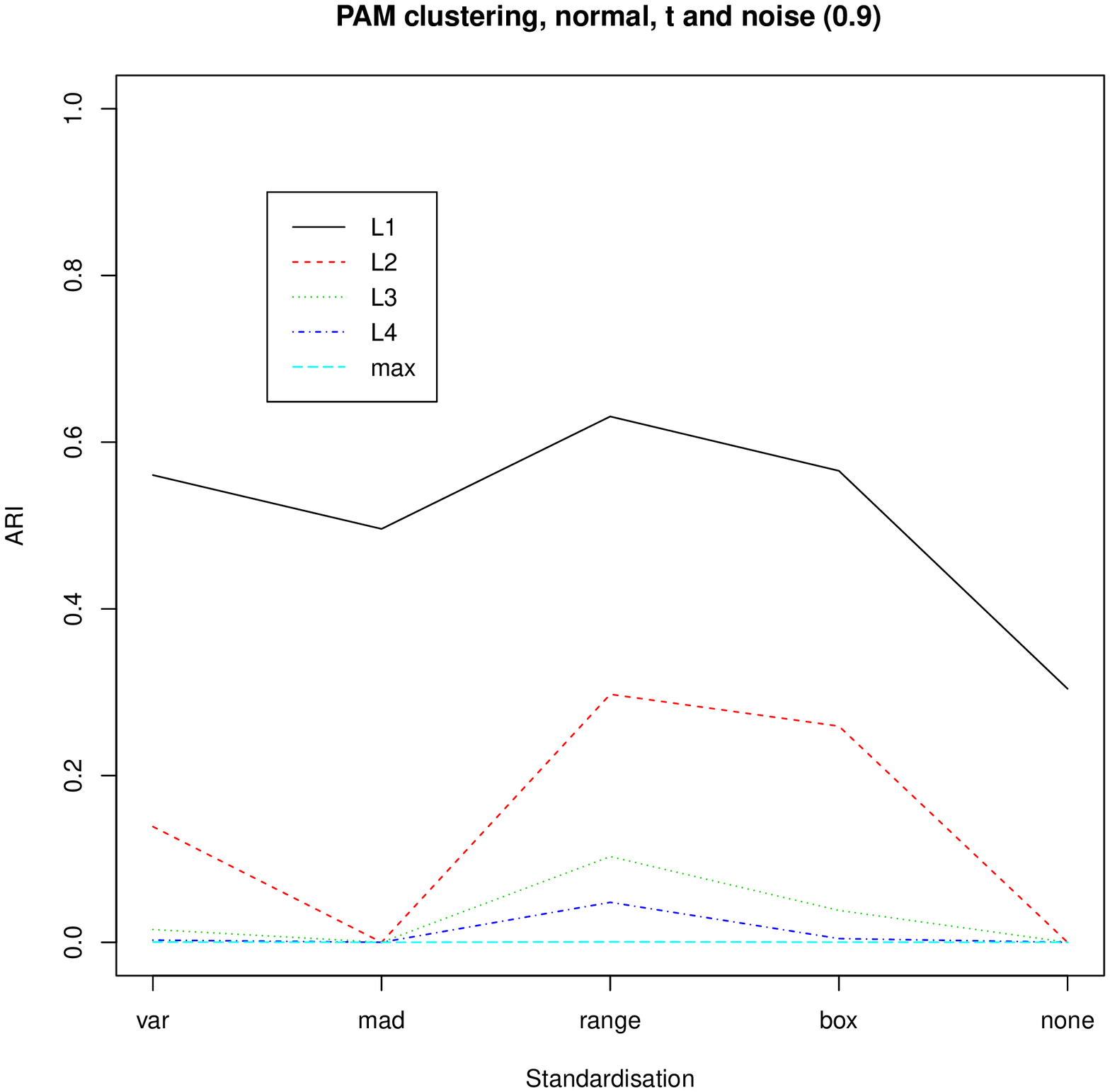}
\includegraphics[width=0.48\textwidth]{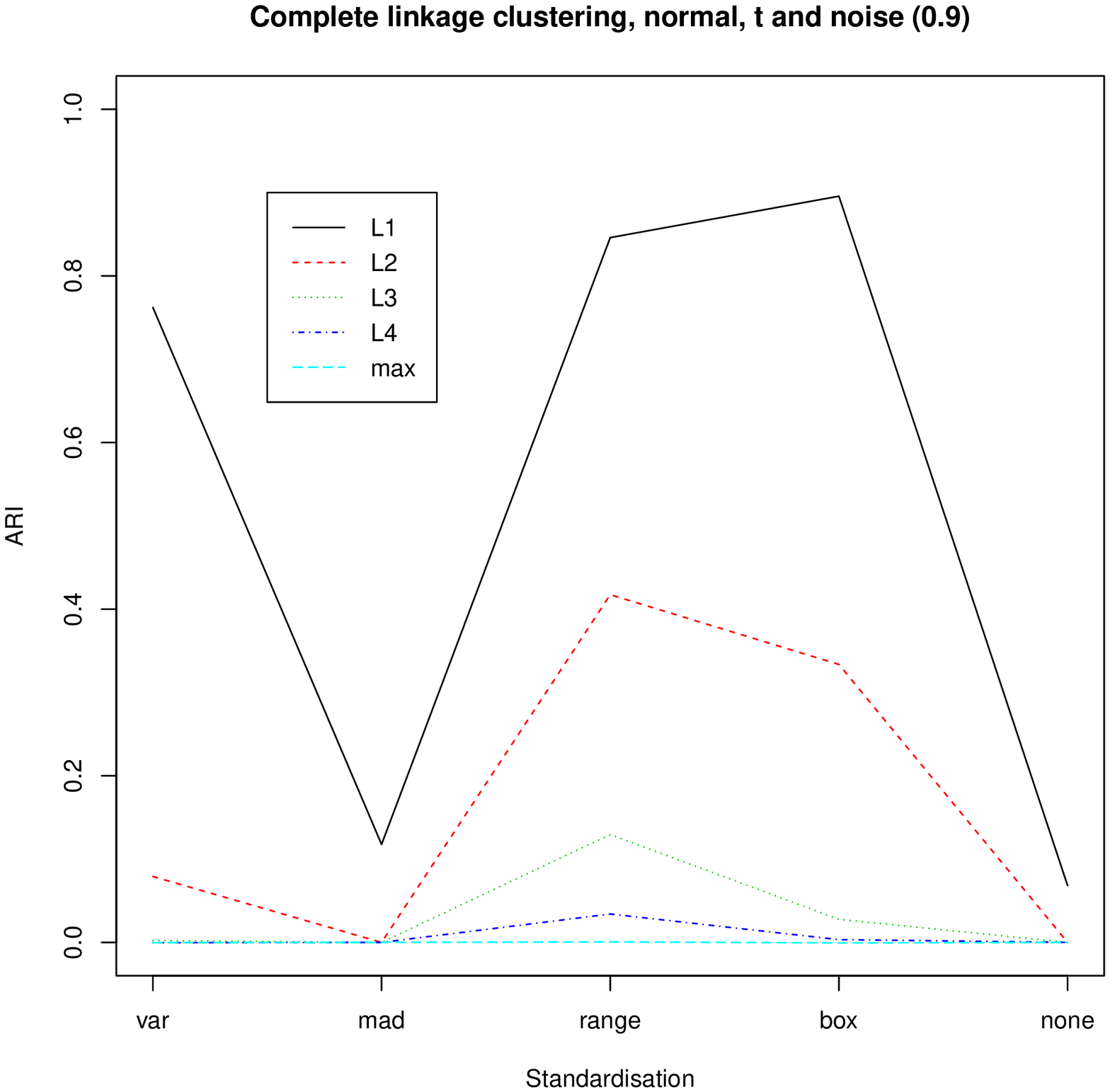}
\includegraphics[width=0.48\textwidth]{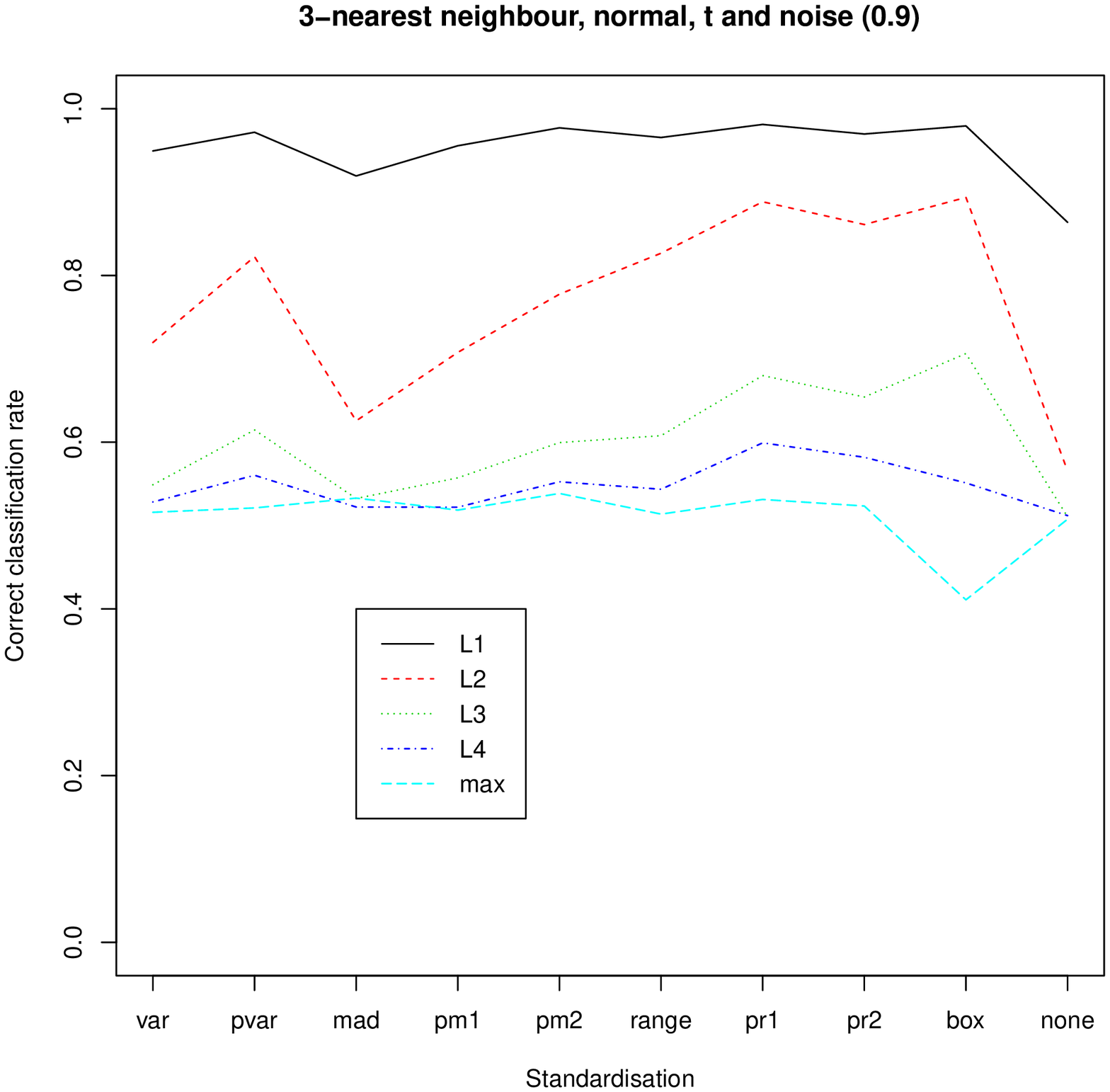}
\end{center}
%
%
\caption{Results from the normal, t, and noise (0.9) setup, adjusted Rand index (ARI) from PAM and complete linkage, and misclassification rates from 3-nearest neighbours.}
\label{fnormaltnoise09}       
\end{figure}

I ran some simulations in order to compare all combinations of standardisation and aggregation on some clustering and supervised classification problems. 

\subsection{Setups}
The scope of these simulations is somewhat restricted. In all cases training data were generated with two classes of 50 observations each (i.e., $n=100$) and $p=2000$ dimensions. For supervised classification, test data was generated according to the same specifications. All variables were independent. Variables were generated according to either Gaussian or $t_2$-distributions within classes (the latter in order to generate string outliers). The mean differences between the two classes were generated randomly according to a uniform distribution, as were the standard deviations in case of a Gaussian distribution; a $t_2$-random variable was just multiplied by the value corresponding to a Gaussian standard deviation. Standard deviations were allowed to differ between classes; in fact they were independently drawn. With probability $p_t$, a variable was chosen to be $t_2$-distributed, otherwise Gaussian. With probability $p_n$, a variable was ``noise'', i.e., there was no distributional difference between the classes. Draws from $p_t$ and $p_n$ were independent, i.e., both noise and informative variables could be t-distributed. 

There were five different setups:
\begin{description}
\item[Simple normal] $p_t=p_n=0$ (all distributions Gaussian and with mean differences), all mean differences 0.1, standard deviations in $[0.5,1.5]$.
\item[Simple normal (0.99)] $p_t=0$ (all Gaussian) but $p_n=0.99$, much noise and clearly distinguishable classes only on 1\% of the variables. All mean differences 12, standard deviations in $[0.5,2]$.
\item[Normal, t, and noise (0.1)] $p_t=p_n=0.1$, mean differences in $[0,0.3]$,  standard deviations in $[0.5,10]$. Weak information on many variables, strongly varying within-class variation, outliers in a few variables.
\item[Normal, t, and noise (0.5)] $p_t=p_n=0.5$, mean differences in $[0,2]$,  standard deviations in $[0.5,10]$. Half of the variables with mean information, half of the variables potentially contaminated with outlier, strongly varying within-class variation.
\item[Normal, t, and noise (0.9)] $p_t=p_n=0.9$, mean differences in $[0,10]$,  standard deviations in $[0.5,10]$. Only 10\% of the variables with mean information, 90\% of the variables potentially contaminated with outlier, strongly varying within-class variation.
\end{description}
For clustering, PAM, average and complete linkage were run, all with number of clusters known as 2. Results were compared with the true clustering using the adjusted Rand index (\cite{HubAra85}). Results for average linkage are not shown because it always performed worse than complete linkage, probably mostly due to the fact that cutting the average linkage hierarchy at 2 clusters would very often produce a one-point cluster (single linkage would be even worse in this respect). For supervised classification, a 3-nearest neighbour classifier was chosen, and the misclassification rate on the test data was computed. There were 100 replicates for each setup.

\subsection{Results}
Results are shown in Figures \ref{fsimplenormal}-\ref{fnormaltnoise09}. These are interaction (line) plots showing the mean results of the different standardisation and aggregation methods. I had a look at boxplots as well; it seems that differences that are hardly visible in these plots are in fact insignificant, taking into account random variation (which cannot be assessed from the interaction plots alone), and things that seem clear are also 
significant. 
``pvar'' stands for pooled variance, ``pm1'' and ``pr1'' stand for weights-based pooled MAD and range, respectively, and ``pm2'' and ``pr2'' stand for shift-based pooled MAD and range, respectively. 

The clearest finding is that $L_1$-aggregation is the best in almost all respects, often with a big distance to the others. It is hardly ever beaten; only for PAM and complete linkage with range standardisation  clustering in the simple normal (0.99) setup  (Figure \ref{fsimplenormal099}) and PAM clustering in the simple normal setup (Figure \ref{fsimplenormal}) some others are slightly better. Actually $L_1$-aggregation delivers a good number of perfect results (i.e., ARI 1 or misclassification rate 0). This is in line with \cite{HAK00}, who state that ``the $L_1$-metric is the only metric for which the absolute difference between nearest and farthest neighbor increases with the dimensionality.''

Results for $L_2$ are surprisingly mixed, given its popularity and that it is associated with the Gaussian distribution present in all simulations. It is in second position in most respects but performs worse for PAM clustering (normal, t, and noise (0.1 and 0.5), simple normal (0.1)), where $L_4$ holds the second and occasionally even the first position. $L_3$ and $L_4$ generally performed better with PAM clustering than with omplete linkage and 3-nearest neighbour. The reason for this is that $L_ 3$ and $L_4$ are
dominated by the variables on which the largest distances occur. This means that very
large within-class distances can occur, which is bad for complete linkage’s chance of
recovering the true clusters, and also bad for the nearest neighbour classification of
most observations. Still PAM can find cluster centroid objects that are only extreme
on very few if any variables and will therefore be close to most of not all observations
within the same class. On the other hand, almost generally, it seems more favourable
to aggregate information from all variables with large distances as $L_3$ and $L_4$ do than
to only look at the maximum. 

Regarding the standardisation methods, results are mixed. The boxplot transformation performs overall very well and often best, but the simple normal (0.99) setup (Figure \ref{fsimplenormal099}) with a few variables holding strong information and lots of noise shows its weakness. In such a case, for clustering range standardisation works better, and for supervised classification using pooling is better. A higher noise percentage is better handled by range standardisation, particularly in clustering; the standard deviation,
MAD and boxplot transformation can more easily downweight the variables that
hold the class-separating information.  The simple normal (0.99) setup is also the only one in which good results can be achieved without standardisation, because here the variance is informative about a variable's information content. Otherwise standardisation is clearly favourable (which it will more or less always be for variables
that do not have comparable measurement units). 

For supervised classification, the advantages of pooling can clearly be seen for the higher noise proportions (although the boxplot transformation does an excellent job for normal, t, and noise (0.9)); for noise probabilities 0.1 and 0.5 the picture is less clear. In the latter case the MAD is not worse than its pooled versions, and the two versions of pooling are quite different. Weights-based pooling is better for the range, and shift-based pooling is better for the MAD. In these setups the mean differences between the classes are dominated by their variances; pooling is much better only where much of the overall variance, MAD or range is caused by large between-class differences. On the other hand, with more noise (0.9, 0.99) and larger between-class differences on the informative variables, MAD-standardisation does not do well. Despite its popularity variance and even pooled variance standardisation are hardly ever among the best methods.   

A curiosity is that some misclassification percentages, particularly for $L_3, 
L_4$ and maximum aggregation, are clearly worse than 50\%, meaning that the methods do worse than random guessing, e.g. in the right graph of Figure \ref{fsimplenormal}. The reason for this is that with strongly varying within-class variances for a given pair of observations from the same class the largest distance is likely to stem from a variable with large variance, and the expected distance to an observation of the other class with typically smaller vaiance will be smaller (although with even more variables it may be more reliably possible to find many variables that have a variance near the maximum
simulated one simultaneously in both classes, so that the maximum distance can be
dominated by the mean difference between the classes again, among those variables
with near maximum variance in both classes).

\section{Conclusion}
\label{sconclusion}
Distance-based methods seem to be underused for high dimensional data with low sample sizes, despite their computational advantage in such settings. This is partly due to undesirable features that some distances, particularly Mahalanobis and Euclidean, are known to have in high dimensions. This work shows that the $L_1$-distance in particular has a lot of largely unexplored potential for such tasks, and that further improvement can be achieved by using intelligent standardisation. The boxplot transformation proposed here performed very well in the simulations expect where there was a strong contrast between many noise variables and few variables with strongly separated classes. In such situations dimension reduction techniques will be better than impartially aggregated distances anyway. For supervised classification it is often better to pool within-class scale statistics for standardisation, although this does not seem necessary if the difference between class means does not contribute much to the overall variance. 

The simulations presented here are of limited scope. Dependence between variables should be explored, as should larger numbers of classes and varying class sizes. Standardisation methods based on the central half of the observations such as MAD and boxplot transformation may suffer in presence of small classes that are well separated from the rest of the data on individual variables. 

%
%
%

\end{document}